\documentclass[10pt]{article}

\usepackage{blois}
\usepackage{graphicx} \graphicspath{{figures/}}
\usepackage{amssymb}
\usepackage{url}

\begin{document}

\vspace*{4cm}

\title{SEARCHES FOR THE HIGGS BOSON AT LHC}

\author{
  M.~DELMASTRO\\
  \vspace{1mm}
  {\footnotesize \emph{on behalf of the ATLAS and CMS collaborations}}
}

\address{
  \vspace{2mm}
  European Laboratory for Particle Physics (CERN)\\
  CH-1211 Geneva 23, Switzerland
  \vspace{2mm}
}

\maketitle

\abstracts{The search strategy for the Standard Model Higgs boson at
  the Large Hadron Collider is reviewed, with a particular emphasis on
  its potential observation by the ATLAS and CMS detectors in the
  $\gamma\gamma$, $\tau^+\tau^-$, $ZZ^{*}$ and $WW^{*}$ final
  states. The combined Higgs discovery potential of ATLAS and CMS is
  discussed, as well as the expected exclusion limits on the
  production rate times the branching ratio as a function of the Higgs
  mass and the collected luminosity.}

\newcommand{\GeV} {\hbox{${\rm GeV}/c^{2}$}}
\newcommand{\TeV} {\hbox{${\rm TeV}$}}
\newcommand{\mH}  {\hbox{$m_{H}$}}

\section{Introduction}

The main goal of the Large Hadron Collider\,\cite{LHC} (LHC) is to
shed light on the mechanism responsible for the electroweak symmetry
breaking. In the context of the Standard Model\,\cite{sm_all} (SM)
this is ensured by the Brout-Englert-Higgs mechanism\,\cite{higgs_all}
that, by assuming the existence of one doublet of scalar fields, gives
also rise to an additional scalar particle known as the Higgs
boson. The mass of the Higgs boson \mH{} is not predicted by the
theory\,\cite{Djouadi20081-short}, but direct experimental searches
have set a lower limit\,\cite{LEP2-short} to $\mH>114.4\,\GeV$, and
have recently excluded\,\cite{arXiv:0903.4001-short} the range between
160 and 170\,\GeV{} at the 95\% confidence level (CL). The global fit
of precision electroweak data\,\cite{Gfitter-2009-short} including the
LEP--2 data\,\cite{LEP2-short} indicates a preferred mass of
$\mH=116.4_{-1.3}^{+15.6}\,\GeV$ with an upper bound of 191\,\GeV{},
if the Tevatron direct limits\,\cite{arXiv:0903.4001-short} are not
included \cite{LEP-EW-WG}.

The two LHC general-purpose detectors
ATLAS\,\cite{1748-0221-3-08-S08003} and
CMS\,\cite{1748-0221-3-08-S08004} are designed to search for the SM
Higgs boson over a wide mass range\,\cite{ATLAS-EP,CMT-TDR-2}, from
the LEP exclusion limit up to several hundreds of \GeV{}. An overview
of the the search strategies over the whole mass range is presented
here, as well as the sensitivities of the detectors with different
integrated luminosities at $\sqrt{s} = 14\,\TeV$, the nominal LHC
center-of-mass energy for the proton--proton collisions.

\section{The Standard Model Higgs boson at the LHC}

\begin{figure}[ht]
  \begin{minipage}[t]{0.49\linewidth}
    \centering
    \includegraphics[height=.6\linewidth]{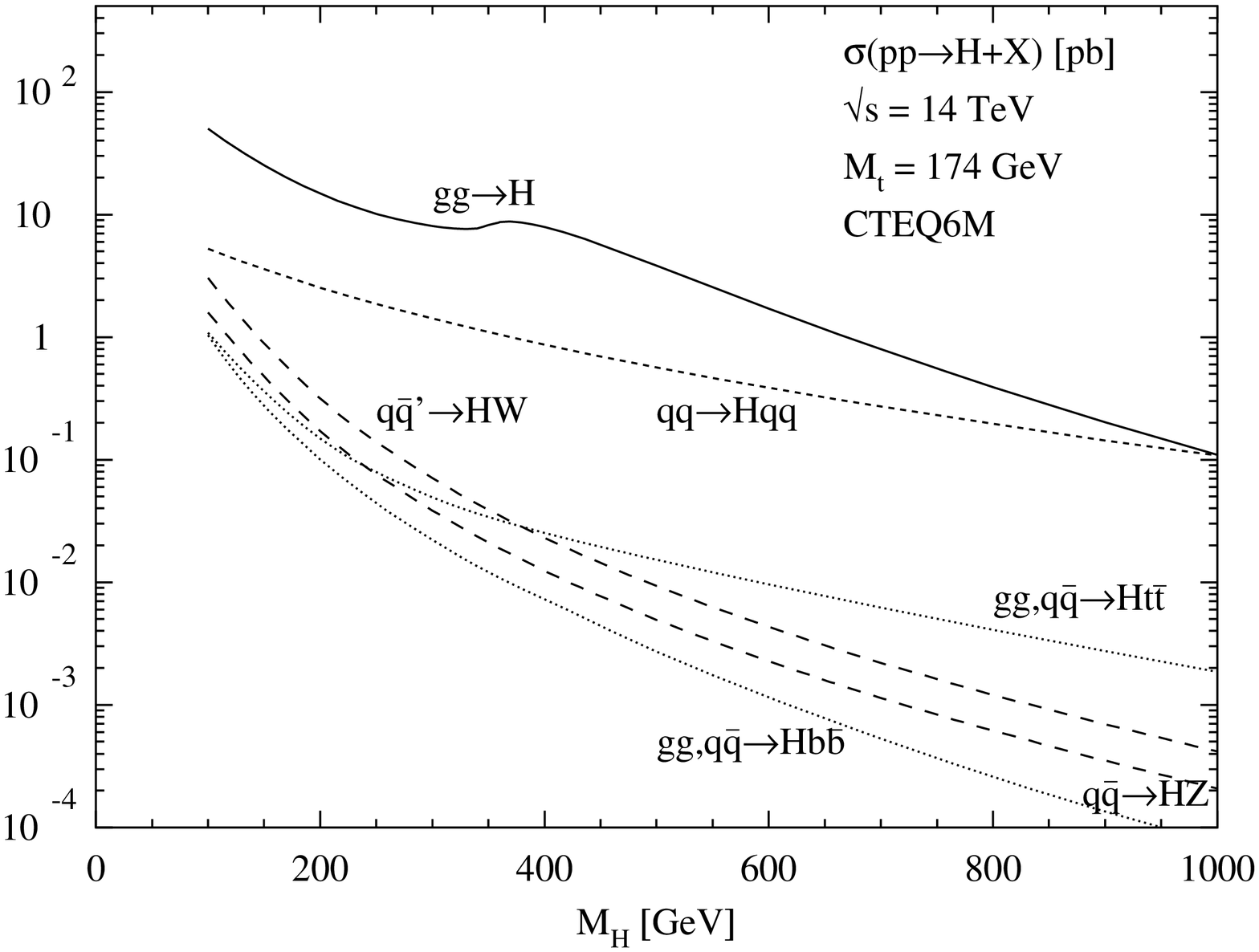}
    \caption{SM Higgs production cross-sections at the LHC for
      $\sqrt{s} = 14 \TeV$ as a function of the Higgs mass.}
    \label{fig:HiggsProd}
  \end{minipage}
  \hspace{0.02\linewidth}
  \begin{minipage}[t]{0.49\linewidth}
    \centering
    \includegraphics[height=.6\linewidth]{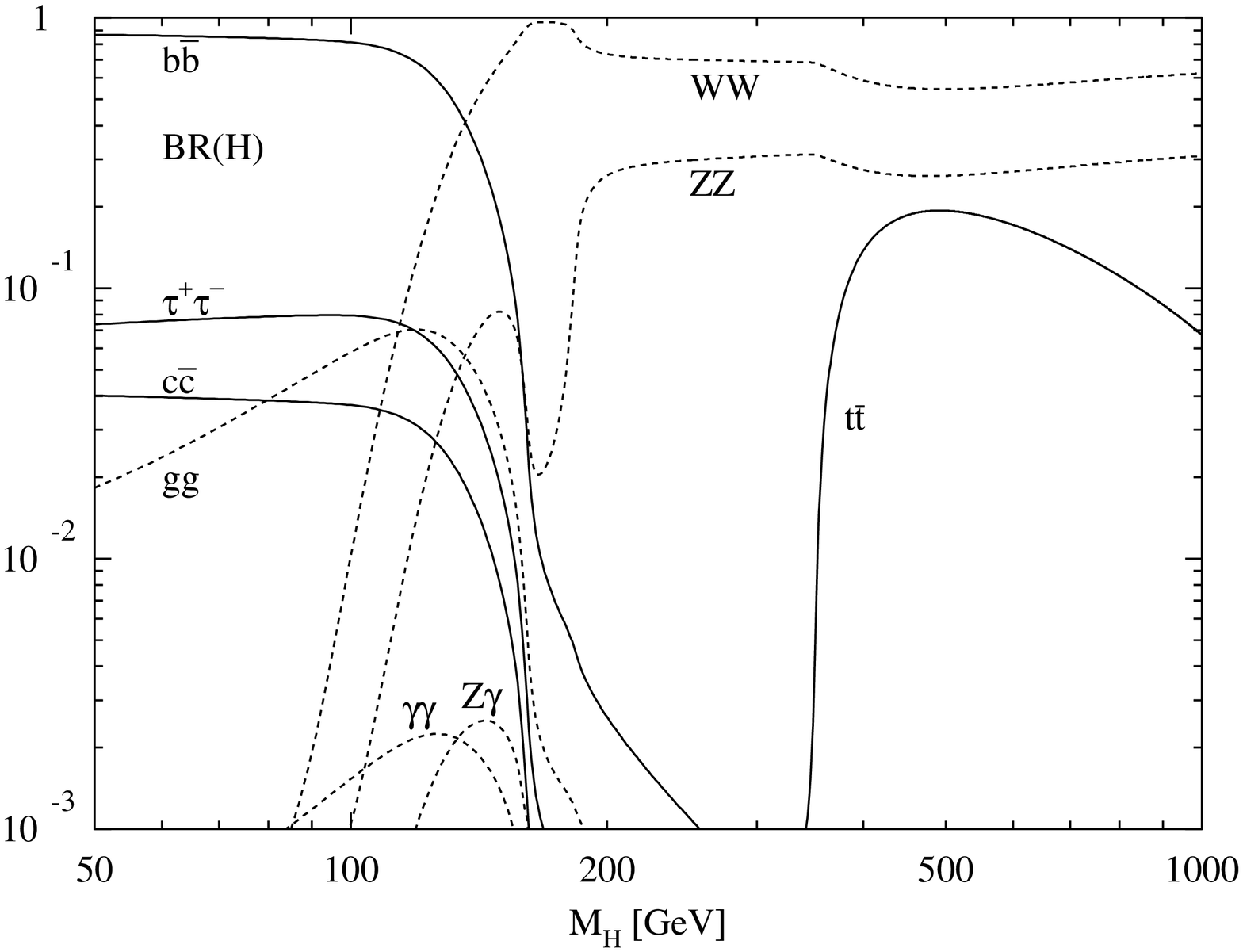}
    \caption{Branching ratios for SM Higgs decays as a function of the
      Higgs mass.}
    \label{fig:HiggsDecays}
  \end{minipage}
\end{figure}

The dominant production mechanism of the SM Higgs boson in the
proton--proton collisions provided by the LHC is the gluon
fusion\,\cite{Djouadi20081-short} (Fig.~\ref{fig:HiggsProd}), where at
leading order the production is mediated by an heavy quark loop. The
next-to-leading (NLO) order cross section for this process accounts to
about 37.6\,pb for $\mH=120\,\GeV$. The following production mechanism
is trough Vector Boson Fusion (VBF)\,\cite{Djouadi20081-short};
despite being approximately one order of magnitude smaller -- the NLO
cross section for this process accounts to about 3.19\,pb for
$\mH=120\GeV$ -- this mechanism is nevertheless interesting for the
particular topologies of its final states. 

The productions of the Higgs boson in association with a $Z$ or $W$
vector boson or with a $t\bar{t}$ pair have much smaller cross
sections: they are used by ATLAS and CMS to improve their sensitivity
further more, or to study very specific final states, but they will
not be discussed here.

\section{Higgs discovery final states}

The most promising Higgs decay modes to search for the particle in a
given mass range are selected both by the Higgs branching
ratios\,\cite{Djouadi20081-short} (see Fig.~\ref{fig:HiggsDecays}) and
by the relative level of background for those particular decays in
that mass range.

In the low mass region ($m_H\lesssim 135 \,\GeV$) the $b\bar{b}$ final
state accounts for about 81\% of the decays, but it is strongly
disfavored because of the very high QCD background and the small $p_T$
of the decay particles. In this mass region the
$H\rightarrow\gamma\gamma$ decay, despite the tiny branching ratio
($\sim$\,0.2\% for $\mH=120\,\GeV$), provides instead a clear
signature of high $p_T$ photons, and represents the most promising
search channel (see Sec.~\ref{sec:Hgg}). The decay in $\tau^+\tau^-$
pairs accounts only for about 8\% of the decays at low mass, and
potentially suffers of limitations similar to the ones affecting the
$b\bar{b}$ decay mode; on the other hand the use of the particular
final state topology provided by the VBF production mode promotes this
decay channel to be an important support to increase the sensitivity
in the low mass region (see Sec.~\ref{sec:Htautau})

For larger masses ($\mH\gtrsim 130\,\GeV$) the decay in a pair of $Z$
or $W$ bosons becomes accessible, with at least one of them
on-shell. Thanks to its kinematics providing a very clear signature,
the decay of the Higgs boson into four leptons mediated by two $Z$
bosons ($H\rightarrow ZZ^{(*)}\rightarrow 4l$) represents the search
\emph{golden channel} in this mass range (see Sec.~\ref{sec:H4l}),
except for the region around $\mH\simeq 2 m_{W}$ where the decay in
two $W$ bosons account for about 95\% of the branching ratio, and the
search for $H\rightarrow W W^{(*)} \rightarrow l \nu l \nu$ becomes
the most significant one (see Sec.~\ref{sec:HWW}).

\subsection{$H\rightarrow\gamma\gamma$}
\label{sec:Hgg}

\begin{figure}
  \begin{minipage}[t]{0.49\linewidth}
    \centering
    \includegraphics[height=.6\linewidth]{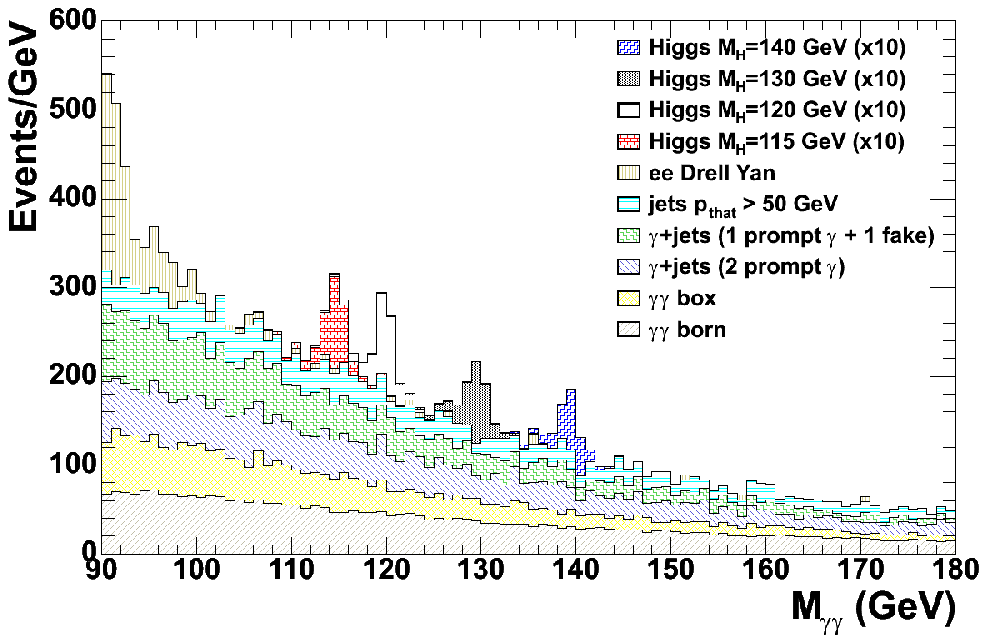}
    \caption{Di-photon invariant mass spectrum after the selection for
      the CMS cut-based analysis. Events are normalized to an
      integrated luminosity of 1 fb$^{-1}$ and the Higgs signal, shown
      for different masses, is scaled by a factor 10.}
    \label{fig:HiggsGammaGammaSpectrumCMS}
  \end{minipage}
  \hspace{0.02\linewidth}
  \begin{minipage}[t]{0.49\linewidth}
    \centering
    \includegraphics[height=.6\linewidth]{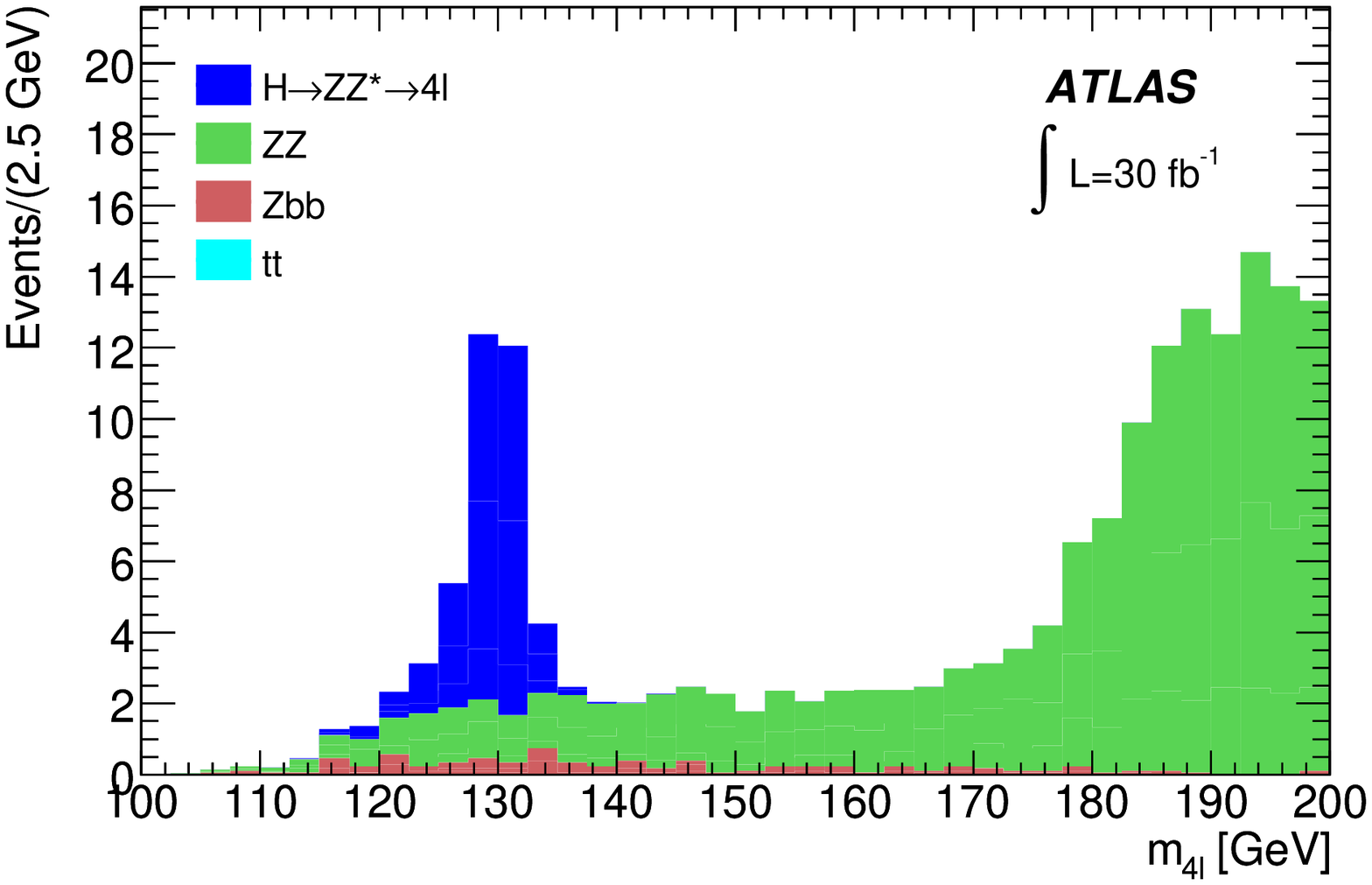}
    \caption{Reconstructed four-lepton mass for signal and background
      processes in ATLAS, in the case of a 130 \GeV{} Higgs boson,
      normalized to a luminosity of 30 fb$^{-1}$.}
    \label{fig:Higgs4LeptonsSpectrumATLAS}
  \end{minipage}
\end{figure}

Two high $p_T$ photons represent the clean signature of this final
state: the kinematic of the event can be fully reconstructed, and the
existence of the Higgs boson would manifest itself as a bump in the
di-photon invariant mass spectrum, sitting on top of the irreducible
background spectrum constituted by genuine photon pairs from
$q\bar{q}\rightarrow\gamma\gamma$, $gg\rightarrow\gamma\gamma$ and
quark bremsstralung (Fig.~\ref{fig:HiggsGammaGammaSpectrumCMS}).

Jet--jet and $\gamma$--jet events where the jets are misidentified as
photons make up the reducible background, that has to be kept as low
as possible with an excellent jet rejection. In order to achieve the
higher sensitivity possible the best invariant mass resolution is
needed: this implies guaranteeing an optimal electromagnetic
calorimetric resolution, an excellent measurement of the interaction
primary vertex, and a good control of the photon conversions.

Both ATLAS and CMS have studied this signal and its background at NLO,
have extended their cut--based analysis's to more
statistical--aggressive
approaches\,\cite{ATLAS-EP,CMS-HiggsGammaGamma-Notes}, and have
reached similar signal sensitivities in this channel. With an
integrated luminosity of 10 (30) fb$^{-1}$ their discovery
significance is about 4\,$\sigma$ (8\,$\sigma$) for $\mH=130\,\GeV$.

\subsection{$H\rightarrow \tau^+\tau^-$}
\label{sec:Htautau}

The search of low mass ($\mH\lesssim 140\,\GeV$) the Higgs boson
decaying in a $\tau^+\tau^-$ pair can be performed exploiting the
particular topology of the events where the Higgs is produces through
VBF. When the Higgs is produced by this mechanism the products of the
$\tau$ decays are accompanied by the presence of two high $p_T$ jets,
that can be used to distinguish the signal. The absence of color flow
between these tag jets introduces a rapidity gap between them: a
central jet veto suppresses then the background.

Because of the small $\tau$ mass with respect to the Higgs mass, the
products of the $\tau$ leptonic or hadronic decays can be considered
approximately collinear: despite the presence of neutrinos in the
final states, the Higgs invariant mass can be fully reconstructed
using the two components of the measured missing transverse energy, if
the Higgs boson has some transverse momentum. The Higgs mass peak
would be partly superimposed to the $Z\rightarrow\tau^+\tau^-$ peak in
the di-$\tau$ invariant mass spectrum. In order to evaluate this
contribution, the similarities between the $Z\rightarrow\mu^+\mu^-$
and $Z\rightarrow\tau^+\tau^-$ processes are used: the measure of the
latter helps to estimate all the detector effects from data, that are
then transposed to the former replacing the measured $\mu$'s with
simulated $\tau$'s.

A 5\,$\sigma$ discovery with this channel alone will require about
30\,fb$^{-1}$ (60\,fb$^{-1}$) for $\mH=115\,\GeV$
($140\,\GeV$)\,\cite{ATLAS-EP,CMS-HiggsTauTau-Notes}.

\subsection{$H\rightarrow Z Z^{(*)} \rightarrow 4l$}
\label{sec:H4l}

The Higgs boson with $\mH\gtrsim 130\,\GeV$ decaying in $ZZ^{(*)}$ is
sought for in the $e^+e^-e^+e^-$, $\mu^+\mu^-\mu^+\mu^-$ and
$e^+e^-\mu^+\mu^-$ channels, taking advantage of the excellent energy
reconstruction of electrons and muons of both the ATLAS and CMS
detectors. Since four leptons are present in all final states, the
reconstruction efficiency plays a crucial role.

The main reducible backgrounds for this channel are $Zb\bar{b}$,
$t\bar{t}$ and fakes, that can be strongly reduced requiring lepton
isolation and imposing selections on the impact parameter; requiring
that at least one $Z$ is on-shell helps reducing the background
further more.

Since the full kinematics of the event can be reconstructed, the Higgs
boson would manifest itself as a clear mass peak of the four leptons
invariant mass sitting on top of a smooth continuum due to the
irreducible background of $ZZ^{(*)}$
(Fig.~\ref{fig:Higgs4LeptonsSpectrumATLAS}), for Higgs masses ranging
from 130 to 600\,\GeV{}.

This channel would guarantees 5\,$\sigma$ discovery significance over
this mass range with about 30\,fb$^{-1}$, except in the region around
2~$m_W$\,\cite{ATLAS-EP,CMS-Higgs4leptons-Notes}. Already with the
1\,fb$^{-1}$ this channel could disprove the existence of an Higgs
boson with $\mH > 185\,\GeV$ at 95\%
CL\,\cite{CMS-Higgs4leptons-1fb-1}.

\subsection{$H\rightarrow W W^{(*)} \rightarrow l \nu l \nu$}
\label{sec:HWW}

Unlike the other channels described above, the Higgs mass peak cannot
be reconstructed in the $H\rightarrow WW^{(*)} \rightarrow l \nu l
\nu$ decays, because of the presence of neutrinos in the final state
and the large $W$ mass. The discovery of the Higgs boson in this
channel reduces to a counting experiment, in which the estimate of the
background level from data is therefore crucial. The background is
dominated by $WW$ and $t\bar{t}$ production, followed by $Wt$, $WZ$,
$ZZ$, Drell--Yan and fakes.

The decay leptons from the $W$'s originating from the Higgs boson
would be preferentially emitted in the same direction in the Higgs
rest frame: this introduces a kinematical correlation that can be used
to separate the signal from the background, and especially to build
signal--depleted samples to measure the background level with.

This powerful decay channel would already allow to exclude the
existence of an Higgs boson between 150 and 180\,\GeV{} at 95\% CL
with the first few fb$^{-1}$, while with 10 fb$^{-1}$ a 5\,$\sigma$
discovery could be claimed\,\cite{ATLAS-EP,CMS-HiggsWW-Notes}.

\section{Higgs discovery and exclusion potentials}

\begin{figure}
  \begin{minipage}[t]{0.49\linewidth}
    \centering
    \includegraphics[height=.7\linewidth]{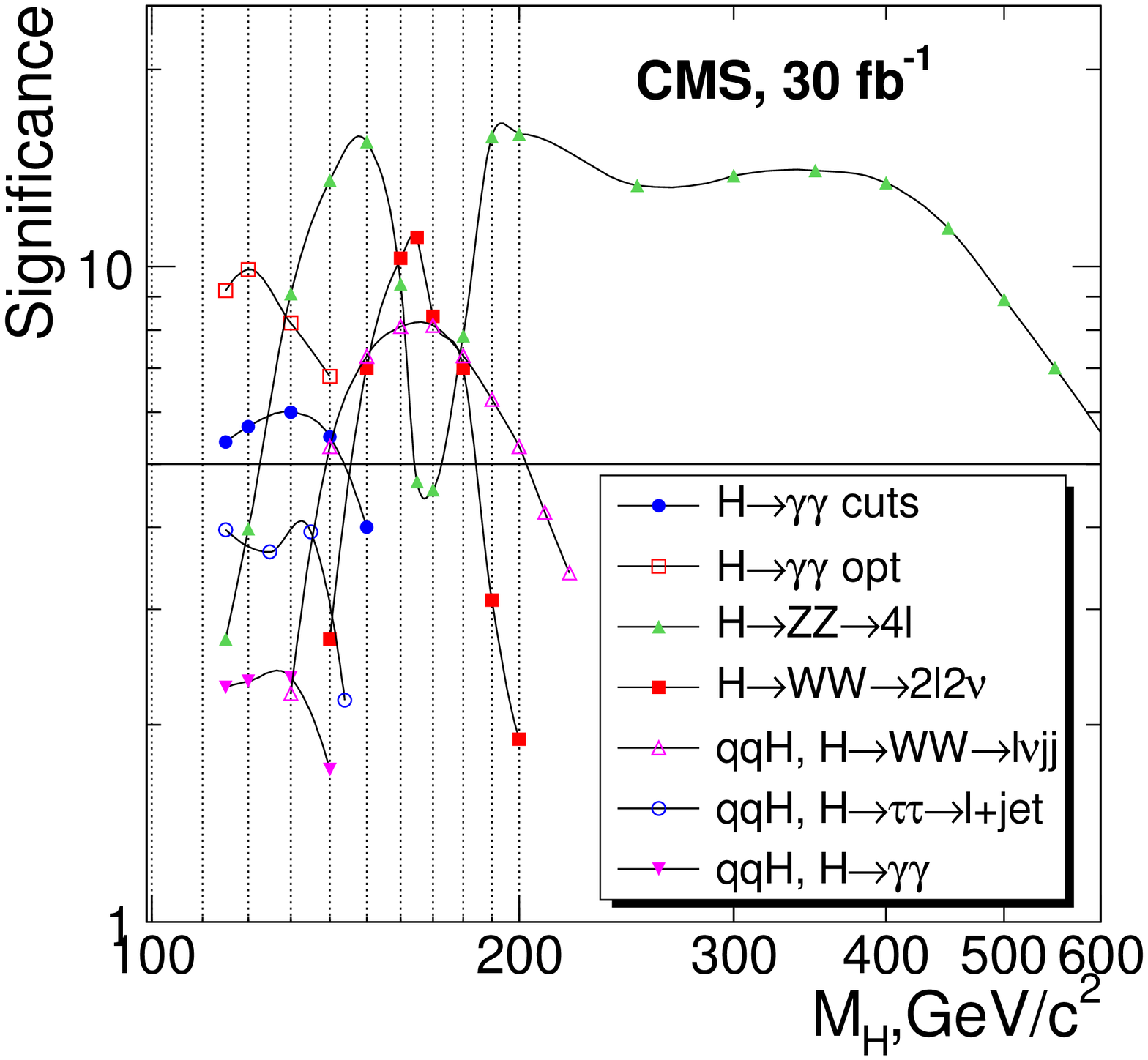}
    \caption{SM Higgs boson signal significance with 30 fb$^{-1}$ in
      CMS, as a function of $\mH$.}
    \label{fig:}
  \end{minipage}
  \hspace{0.02\linewidth}
  \begin{minipage}[t]{0.49\linewidth}
    \centering
    \includegraphics[height=.7\linewidth]{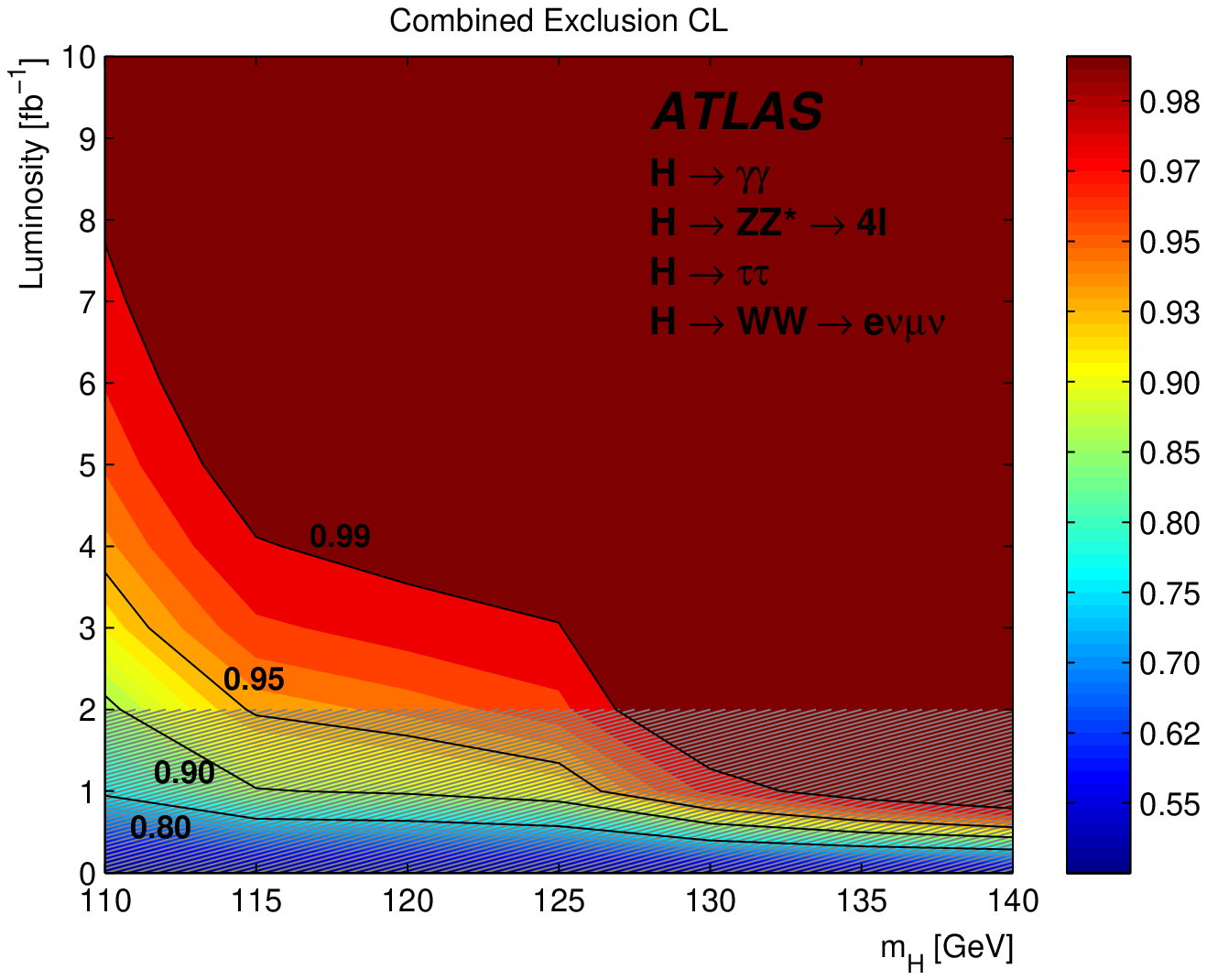}
    \caption{Expected luminosity required by ATLAS to exclude a SM
      Higgs boson with a mass $\mH$ at a confidence level given by the
      corresponding color.}
    \label{fig:HiggsExclusionATLAS}
  \end{minipage}
\end{figure}

The SM Higgs boson discovery reach combining the various final states
described above for an integrated luminosity of 30 fb$^{-1}$ is
presented in Fig.~\ref{fig:} as a function of the Higgs mass for the
CMS experiment\,\cite{CMT-TDR-2}; ATLAS has similar
performances\,\cite{ATLAS-EP}. Both experiments exceed a 5\,$\sigma$
significance over the whole mass range with 30\,fb$^{-1}$ provided by
LHC at $\sqrt{s}=14\,\TeV$.

The combined SM Higgs boson exclusion limits are presented in
Fig.~\ref{fig:HiggsExclusionATLAS} as a function of the Higgs mass and
the integrated luminosity for the ATLAS experiments\,\cite{ATLAS-EP};
CMS has similar performances\,\cite{CMT-TDR-2}. Both experiments will
be able to exclude a SM Higgs boson at 95\% CL over the whole mass
range with the first few fb$^{-1}$ provided by LHC at
$\sqrt{s}=14\,\TeV$.


\section*{References}
\bibliographystyle{unsrt}
{\small
\bibliography{delmastro-blois-2009}
}

\end{document}